\documentclass{iau} 
\usepackage{graphicx}
 \usepackage{amsmath}

\newcommand\HI{$\textrm{H}\scriptstyle\mathrm{I}$}

\def\Htwo{H$_{\,2}$}

\newcommand{\Jyb}{Jy beam$^{-1}$}

\newcommand{\msun}{{${\rm M}_\odot$}}

\newcommand{\forn}{\mbox Fornax~A}

\newcommand{\ngcsix}{\mbox NGC~1316}

\newcommand{\pl}{{\em{Planck}}}
\newcommand{\meer}{{MeerKAT}}

\title[The nuclear activity of \forn] 
{The recurrent nuclear activity of \forn\ and its interaction with the cold gas}

\author[Filippo M. Maccagni]   
{F. M. Maccagni$^1$
P. Serra$^1$,
M. Murgia$^1$,
F. Govoni$^1$,
K. Morokuma-Matsui$^2$,
D. Kleiner$^1$
\thanks{This project has received funding from the European Research Council (ERC) under the European Union’s Horizon 2020 research and innovation programme (grant agreement no. 679627).}
\affiliation{$^1$INAF -- Osservatorio Astronomico di Cagliari, via della Scienza 5, 09047, Selargius (CA), Italy \\ email: {\sc filippo.maccagni@inaf.it} \\[\affilskip]
    $^2$Institute of Astronomy, Graduate School of Science, The University of Tokyo, 2-21-1 Osawa, Mitaka, Tokyo 181-0015, Japan}
}

\pubyear{2020}
\volume{359}  
\jname{Galaxy Evolution and Feedback Across Different Environments}
\editors{A.C. Editor, B.D. Editor \& C.E. Editor, eds.}
\begin{document}

\maketitle

\begin{abstract}
 Sensitive (noise $\sim 16\,\mu$\Jyb), high-resolution ($\sim 10''$) MeerKAT  observations of \forn\ show that its giant lobes have a double-shell morphology, where dense filaments are embedded in a diffuse and extended cocoon, while the central radio jets are confined within the host galaxy. The spectral radio properties of the lobes and jets of \forn\ reveal that its nuclear activity is rapidly flickering. Multiple episodes of nuclear activity must have formed the radio lobes, for which the last stopped $12$ Myr ago. More recently ($\sim 3$ Myr ago), a less powerful and short ($\lesssim 1$ Myr) phase of nuclear activity generated the central jets. The distribution and kinematics of the neutral and molecular gas in the centre give insights on the interaction between the recurrent nuclear activity and the surrounding interstellar medium.
\end{abstract}

\firstsection 
\section{Introduction}

The energy released by Active Galactic Nuclei (AGNs) into the surrounding interstellar medium (ISM) through radiation and/or relativistic jets of radio plasma can drastically change the fate of its host galaxy by removing or displacing the gas in the galaxy and preventing it from cooling to form new stars (e.g.\,\cite[Fabian 2012]{fabian2012}). This mechanism is commonly referred to as `AGN feedback'. Numerical simulations of galaxy evolution indicate that only multiple phases of nuclear activity, and therefore recurring episodes of AGN feedback, may prevent the hot circumgalactic gas from cooling back onto the galaxy, and explain the rapid quenching of star formation in early-type galaxies~(\cite[Werner et al. 2019]{werner2019}). 

The radio emission of AGNs allows us to measure the duty cycle of the nuclear activity. In particular, the steepening of the radio spectrum is often interpreted as radiative ageing of the electron population in the relativistic plasma~(e.g.,~\cite[Murgia et al. 1999, Harwood et al. 2013, Kolokythas et al. 2015]{murgia1999,harwood2013,kolokythas2015}).

\begin{figure}[tbh]
        \begin{center}
                \includegraphics[trim = 0 0 0 0, width=0.44\textwidth]{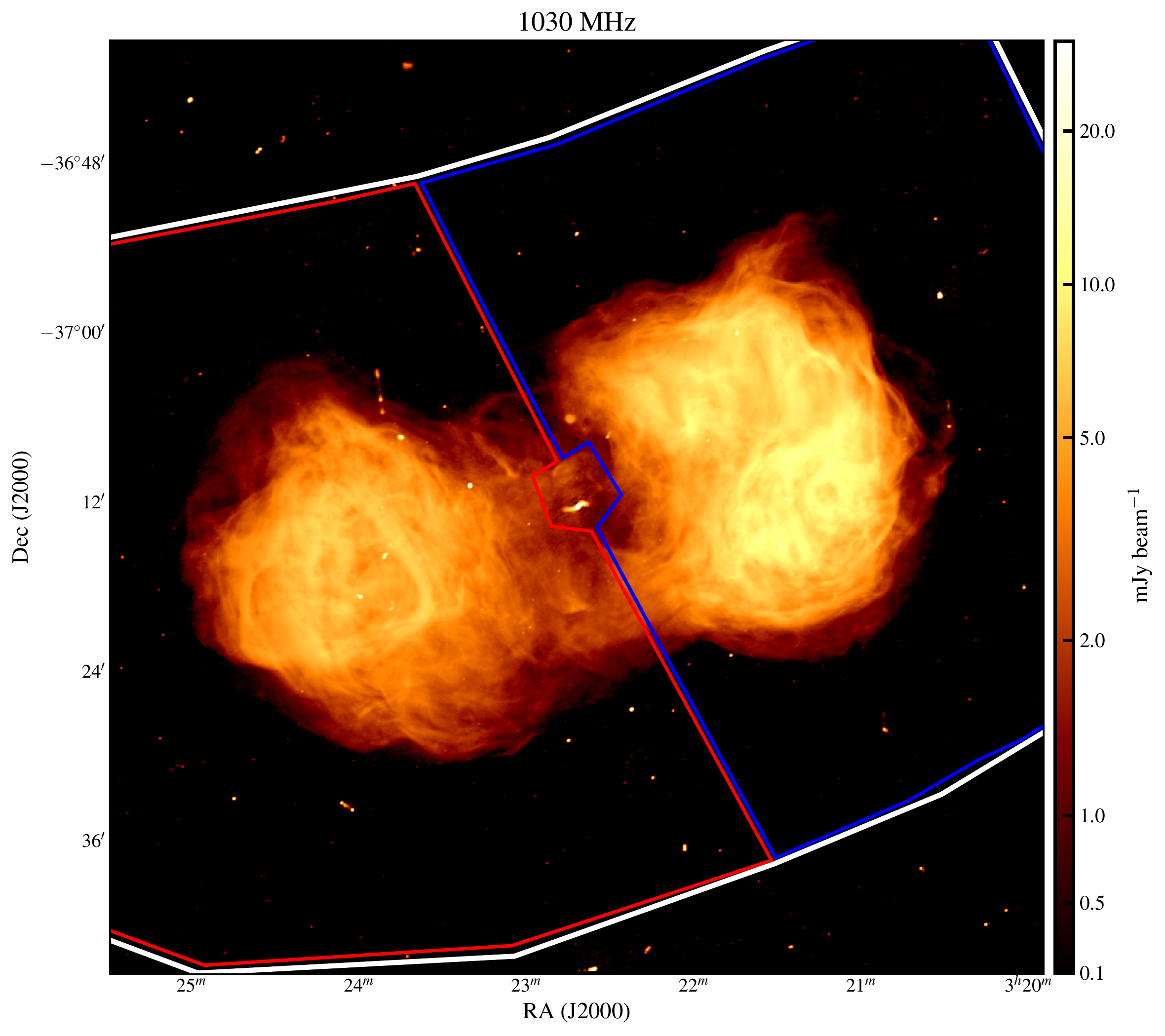}
                \includegraphics[width=0.44\textwidth]{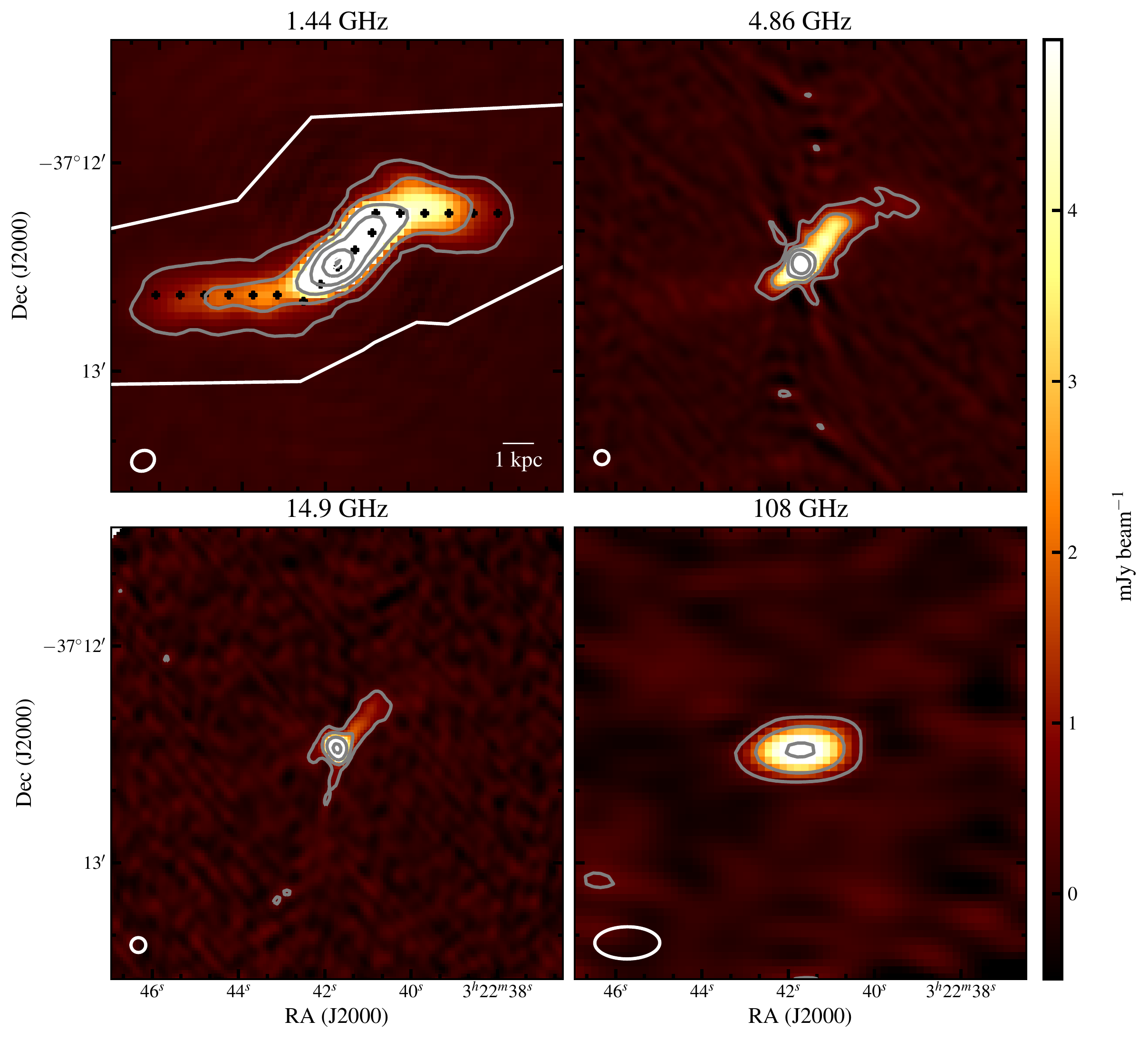}
                \caption{{\em Left panel}: \forn\ seen by MeerKAT at 1.03 GHz. The red and blue contours mark the region where we measure flux density of the east and west lobes, respectively. The synthesised beam of the image is $11.2''\times9.1''$. {\em Right panel}: Central emission of \forn\ seen at 1.44 GHz by \meer\ ({\em top left}), at 4.86 GHz ({\em top right}), 14.9 GHz ({\em bottom left}) by the VLA and at 108 GHz ({\em bottom right}) by ALMA. The PSF of the images is shown in white. (\cite[Maccagni et al. 2020]{maccagni2020}) }
                \label{fig:fornaxAFlRegions} 
        \end{center}
\end{figure}

In this proceeding, we summarize the study of the radio spectrum of the nearby ($D_{\rm L}\sim20$ Mpc) radio galaxy \forn\ to determine the timescale and the duty cycle of its nuclear activity. This analysis is shown in detail in~\cite[Maccagni et al. (2020)]{maccagni2020}.

\forn\ is one of the most fascinating radio sources in the local Universe because of its filamentary extended radio lobes~($\sim 1.1^\circ$, \cite[Fomalont et al. 1989]{fomalont1989}). The \meer~(\cite[Jonas et al. 2016]{jonas2016}) observation (Fig.\,\ref{fig:fornaxAFlRegions}) at $1.03$ GHz shows that the lobes are embedded in a diffuse cocoon, with a `bridge' of synchrotron emission connecting them. In the centre, two radio jets are confined within the host galaxy ($r\lesssim 6$ kpc) and exhibit an s-shaped morphology. Most of the radio emission is produced in the extended lobes. At $1.4$~GHz, their total flux density is $121$ Jy while that of the jets is $\sim300$ mJy.

\forn\ is hosted by the giant early-type galaxy \ngcsix, which is the brightest member of a galaxy group at the outskirts of the Fornax cluster. \ngcsix\ underwent through a major merger that likely brought large amounts of dust, cold molecular gas~(\cite[Horellou et al. 2001, Galametz et al. 2014, Morokuma-Matsui et al. 2019]{horellou2001,galametz2014,morokuma2019}), and neutral hydrogen~(\cite[Horellou et al. 2001, Serra et al. 2019]{horellou2001,serra2019}) into the centre and around the galaxy. This merger occurred $\sim 1$ -- $3$ Gyr ago~(e.g., \cite[Sesto et al. 2018]{sesto2018}), and it may have triggered the nuclear activity of \forn~(e. g.~\cite[McKinley et al. 2015]{mckinley2015}). Nevertheless, large uncertainties remain on the timescale of formation of the radio lobes. Moreover, this past merger event does not properly explain the properties of the central emission, nor the soft X-ray cavities between the lobes and the host galaxy~(\cite[Lanz et al. 2010]{lanz2010}).

The goal of this study is to measure, over a wide range of frequencies, the flux density distribution of the radio lobes and the central emission to characterise the AGN activity history that created them. For the lobes we need wide-field-of-view observations sensitive to their diffuse emission (i.e. good $uv$-coverage on the short baselines), while arc-second resolution is not needed. Hence, between $84$ and $200$~MHz we chose observations Murchison Widefield Array survey~(\cite[Hurley-Walker et al. 2017]{hurleywalker2017}). We use the \meer\ observation to generate images of \forn\ at $1.03$ (Fig.\,\ref{fig:fornaxAFlRegions}, left panel) and $1.44$~GHz. At $1.5$~GHz we chose archival Very Large Array observations (\cite[Fomalont et al. 1989]{fomalont1989}). Between $5.7$ and $6.9$~GHz we use observations from the Sardinia Radio Telescope (\cite[Prandoni et al. 2017]{prandoni2017}). Between $70$~GHz and $217$~GHz, we selected images from the \pl\ foreground maps~(\cite[Planck Collaboration IV, 2018]{planck2018IV}).

To study the central emission we selected observations with arc-second resolution (Fig.\,\ref{fig:fornaxAFlRegions}, right panel): the MeerKAT images at $1.03$ and $1.44$~GHz, archival VLA observations at $4.8$ and $15$~GHz ~(\cite[Geldzahler et al. 1984]{geldzahler1984}), and an observation at $108$~GHz~(\cite[Morokuma-Matsui et al. 2019]{morokuma2019}) taken with the Atacama Large Millimeter and submillimeter Array. The left panel of Fig.\,\ref{fig:tsyncOvertad} shows the spectral flux densities of the lobes and of the central emission we measured using these two samples. Given the morphology of the central emission, we measure its flux density by dividing it into two parts, the central unresolved component (hereafter, the {\em kpc-core}) and the extended component forming the emission (the {\em jets}).

\section{Spectral analysis of the main components of \forn}

In the simplest scenario of AGN activity, the lobes (or jets) are continuously injected with particles ({continuous injection model}, CI). Assuming that radiative energy losses from synchrotron and inverse Compton radiation dominate over expansion losses, the radio spectrum shows a sharp cut-off whose frequency ($\nu_{\rm break}$) depends on the age of the radiation ($t_{s}$, ~\cite[Kardashev et al. 1962]{kardashev1962}).

A more complicated scenario can be the {continuous injection plus turn off model} (CI$_{\rm OFF}$). The injection of high-energy particles from the nucleus starts at $t = 0$ and at the time $t_{\rm CI}$ it is switched off. After that, the {\em off phase} of the AGN begins, and the total age of the radiation is: $t_{\rm s}=t_{\rm CI} + t_{\rm OFF}$ (\cite[Murgia et al. 2011]{murgia2011}). Compared to the CI model, the spectral shape is characterised by a second break-frequency ($\nu_{\rm break,\, high}$), beyond which the radiation spectrum drops exponentially. This frequency depends on the ratio between the dying phase ($t_{\rm OFF}$) and the total age of the source ($\nu_{\rm break,\, high} = \nu_{\rm break} (t_s/t_{\rm OFF})^2$). To determine the best-fit models of the flux density spectrum of the lobes and central emission of Fornax~A, we use the software package {\sc SYNAGE++} (\cite[Murgia et al. 2011]{murgia2011b}).

For both lobes, the spectral flux density shows a sharp cut-off at high frequencies (see Fig.\,\ref{fig:tsyncOvertad}, left panel). This, along with the $\tilde\chi^2$ value of the CI$_{\rm OFF}$ models closer to $1$ than the $\tilde\chi^2$ of the CI models, suggests that the radio spectrum of both radio lobes is best described by the CI$_{\rm OFF}$ model and that currently the radio lobes are not being injected with relativistic particles. 

According to the CI$_{\rm OFF}$ model, the dying-to-total-age ratio is $t_{\rm OFF}/t_s=0.49^{+0.08}_{-0.42}$. Assuming that the magnetic field of the lobes is $2.6\pm 0.3\mu$G~(\cite[Tashiro et al. 2009]{tashiro2009}), the radiative age of the east and west lobes is $t_{s,\,\rm E}=25^{+23}_{-19}$ Myr and $t_{s,\,\rm W}=23^{+20}_{-17}$ Myr, respectively. Likely, in the last $t_{\rm OFF} = 12^{+2}_{-9}$~Myr the lobes have not been replenished with relativistic particles (all parameters derived from the models are shown in Table~6 of~\cite[Maccagni et al. 2020]{maccagni2020}).

The spectral distribution of the kpc-core (in green in Fig.\,\ref{fig:tsyncOvertad}, left panel) is better described by the CI model rather than by the CI$_{\rm OFF}$. The age of the synchrotron emission of the kpc-core is $\sim 1^{+0.3}_{-0.5}$~Myr. By contrast, the jets are better fitted by the CI$_{\rm OFF}$, and do not seem to be currently replenished with energetic particles. Their last active phase seems to have occurred $3^{+7}_{-2}$ Myr ago and to have lasted $\lesssim 1^{+6}_{-0.5}$~Myr.

\firstsection 

\section{The flickering nuclear activity of \forn}
\label{sec:flick}

The properties of the flux density spectrum of the lobes of \forn\ are puzzling when compared to their projected size ($r\sim 200$ kpc). Typically, lobes extending in the IGM for hundreds of kiloparsecs are either the remnant of an old nuclear activity, and show a steep spectrum with low break frequency, or they are currently being injected with relativistic particles, and show a jet or stream of particles connecting the AGN with the lobes. The most remarkable properties of the lobes of \forn\ are the flat spectral shape and high break frequency ($\gtrsim20$ GHz) of their radio emission, and that the nuclear activity that was replenishing the lobes with high-energy particles was short ($\sim 24$~Myr) and has recently stopped ($\sim 12$~Myr ago). The main open question therefore pertains to how these large lobes have formed in such a short time.

\begin{figure}[tbh]
        \begin{center}
                \includegraphics[width=0.44\textwidth]{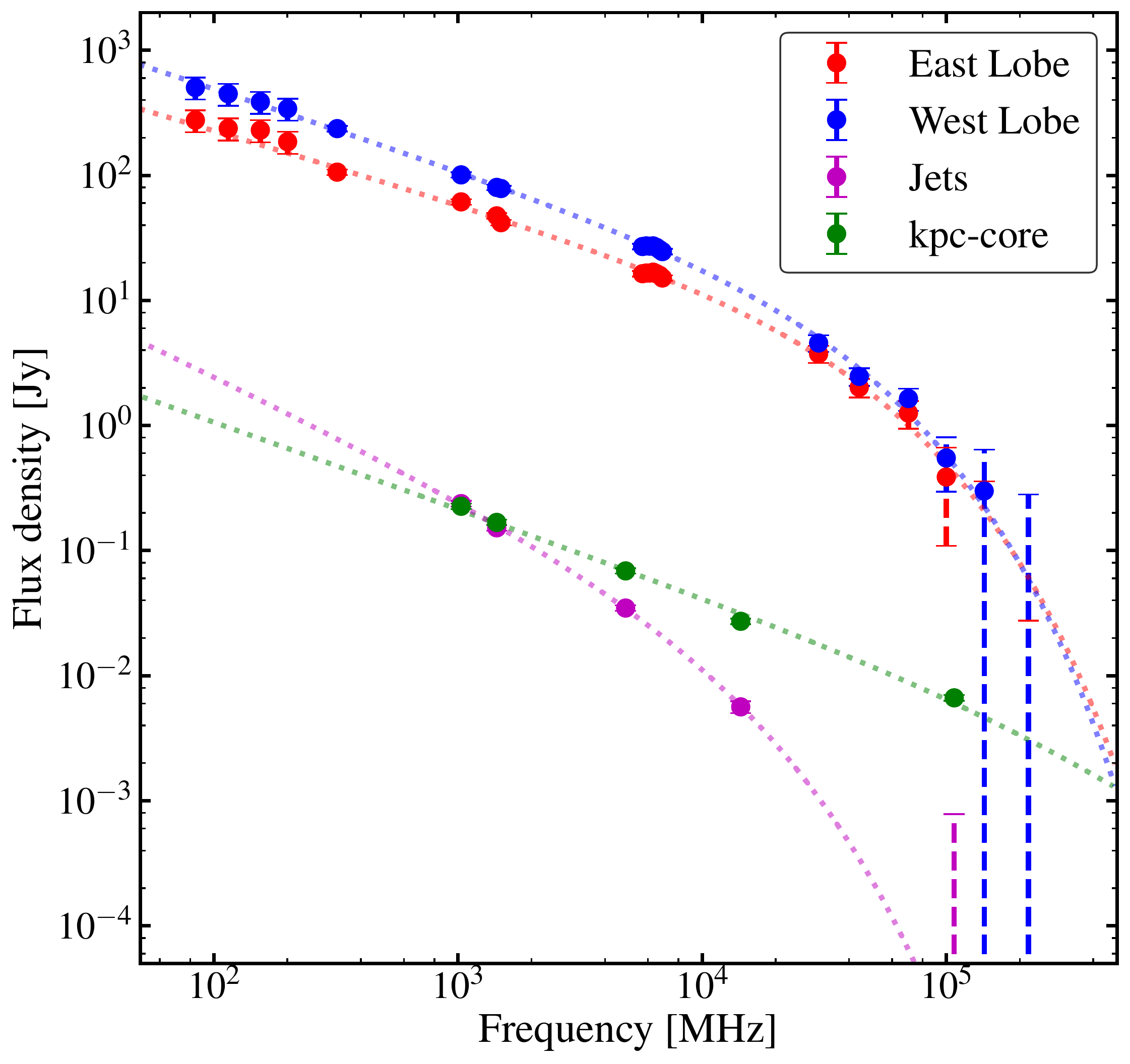}
                \includegraphics[width=0.44\textwidth]{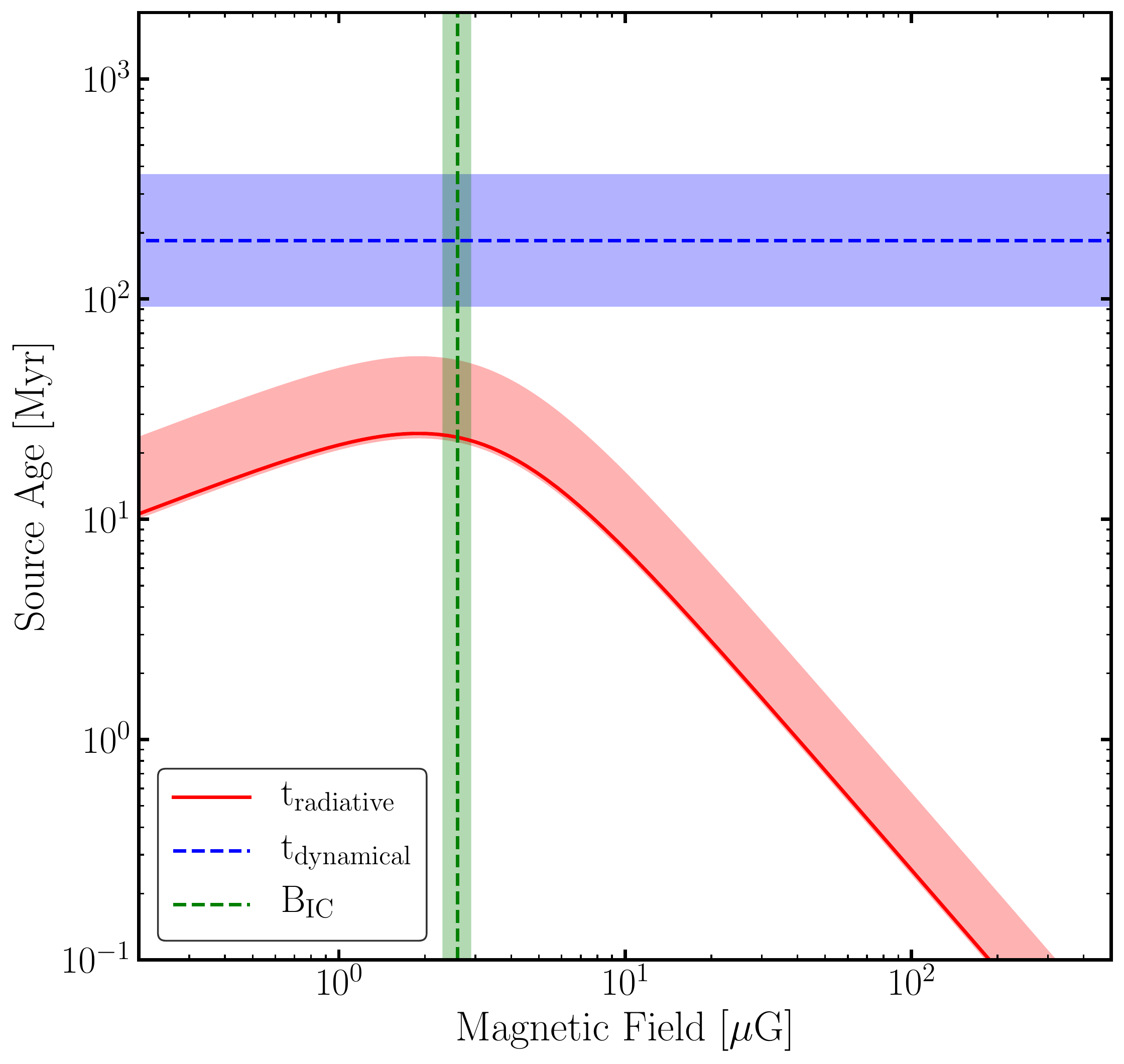}
                \caption{{\em Left panel}: Radio spectrum of the different components of \forn. The east lobe is shown in red, the west lobe in blue. The jets and kpc-core are in magenta and green, respectively. The dashed lines show the CI$_{\rm OFF}$ model of injection that best fits the flux distributions. The spectral shape of the lobes is very different from that of the inner components. {\em Right panel}: Comparison of radiative and dynamical age of \forn\ (red and blue shaded regions) as a function of the magnetic field. The radiative age is derived from the spectral break of the two lobes. Their magnetic field ($B_{IC}\sim2.6\mu$G) is shown in green. (\cite[Maccagni et al. 2020]{maccagni2020}) }
                \label{fig:tsyncOvertad} 
        \end{center}
\end{figure} 

The right panel of Figure~\ref{fig:tsyncOvertad} indicates that the radiative age of the lobes of \forn\ and the dynamical age (assuming transonic expansion) are incompatible. This, along with an axial ratio of the lobes close to 1, disfavours the hypothesis that only a recent episode of activity formed the lobes. If, instead, the lobes, filled with low-energy particles and under-pressured with respect to the surrounding IGM, were already present because of previous activities, a new nuclear phase may rapidly fill them with new high-energy particles which now dominate the radio emission of the source. This scenario would explain the overall flat radio spectrum of \forn\ and its high break-frequency. Two separate AGN outbursts have also been proposed by \cite[Lanz et al. (2010)]{lanz2010} to explain the location of the X-ray cavities relative to the radio lobes of \forn.

Besides the multiple activities that may have formed them, the lobes have been off for $\sim 9$ Myr. More recently ($\sim 3$ Myr ago) the AGN turned on again for a very short phase ($\lesssim 1$ Myr) that formed the central jets. Given the short timescales of the different nuclear activities, \forn\ is likely rapidly flickering between an active phase and a non-active one. The recurrent activity of \forn\ may fit well in the theoretical scenario of AGN evolution whereby the central engine is active for short periods of time ($10^{4-5}$ years), and that these phases repeteadly occur over the total lifetime of the AGN~($10^8$ years; e.g.~\cite[Schawinski et al. 2015, Morganti et al. 2017]{schawinski2015,morganti2017}).

After the major merger, \ngcsix\ went through several accretion events and minor mergers of smaller companions~(\cite[Iodice et al. 2017]{iodice2017}). These numerous interactions may have regulated the switching on and off of the multiple episodes of activity that formed the lobes as we see them now. Merger and interaction events are often invoked to explain the triggering of powerful AGNs~(e. g., \cite[Ramos-Almeida et al. 2012, Sabater et al. 2013]{hopkins2005,ramosalmeida2012,sabater2013}).

\begin{figure}[tbh]
        \begin{center}
                \includegraphics[width=0.44\textwidth]{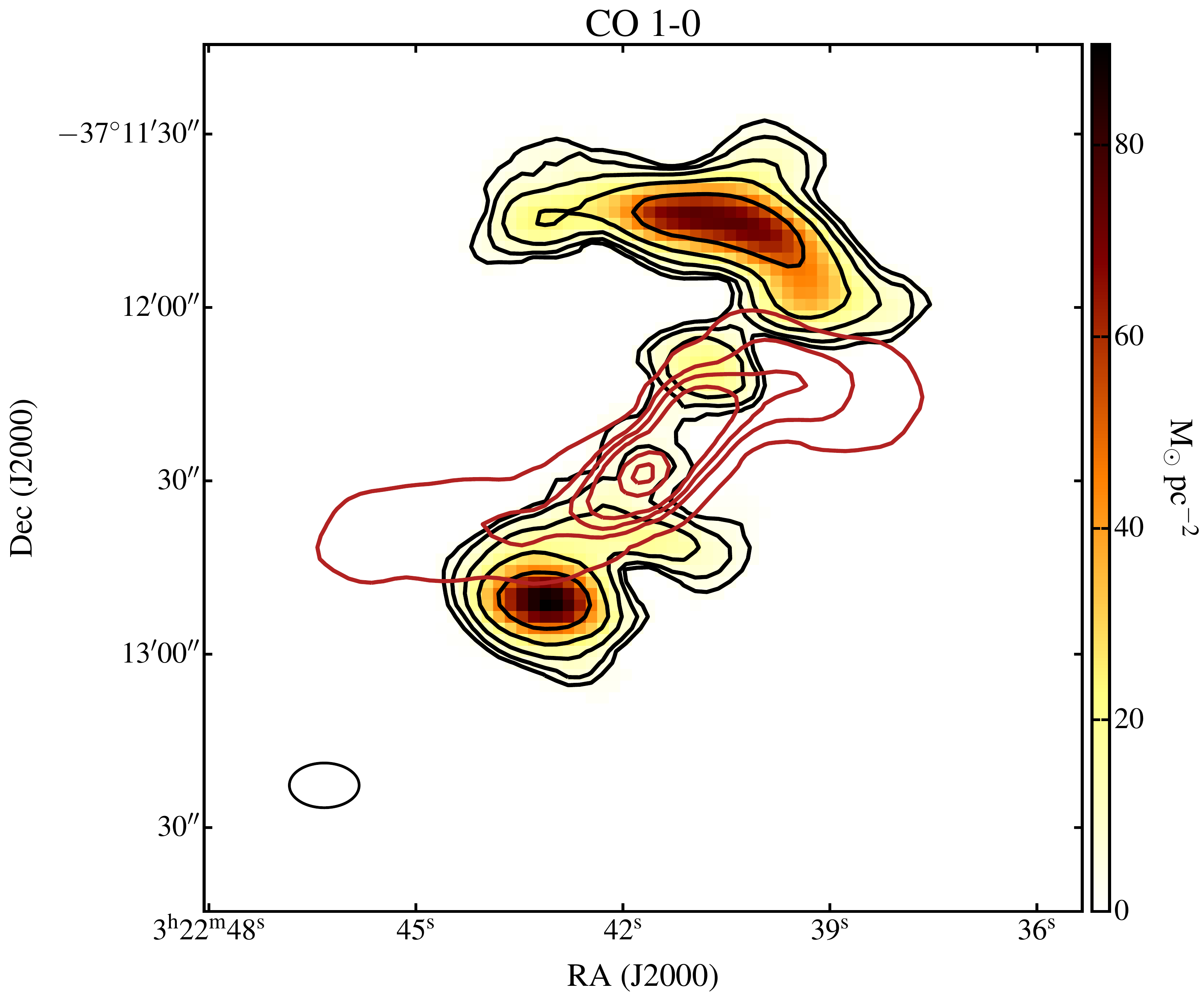}
                \includegraphics[width=0.43\textwidth]{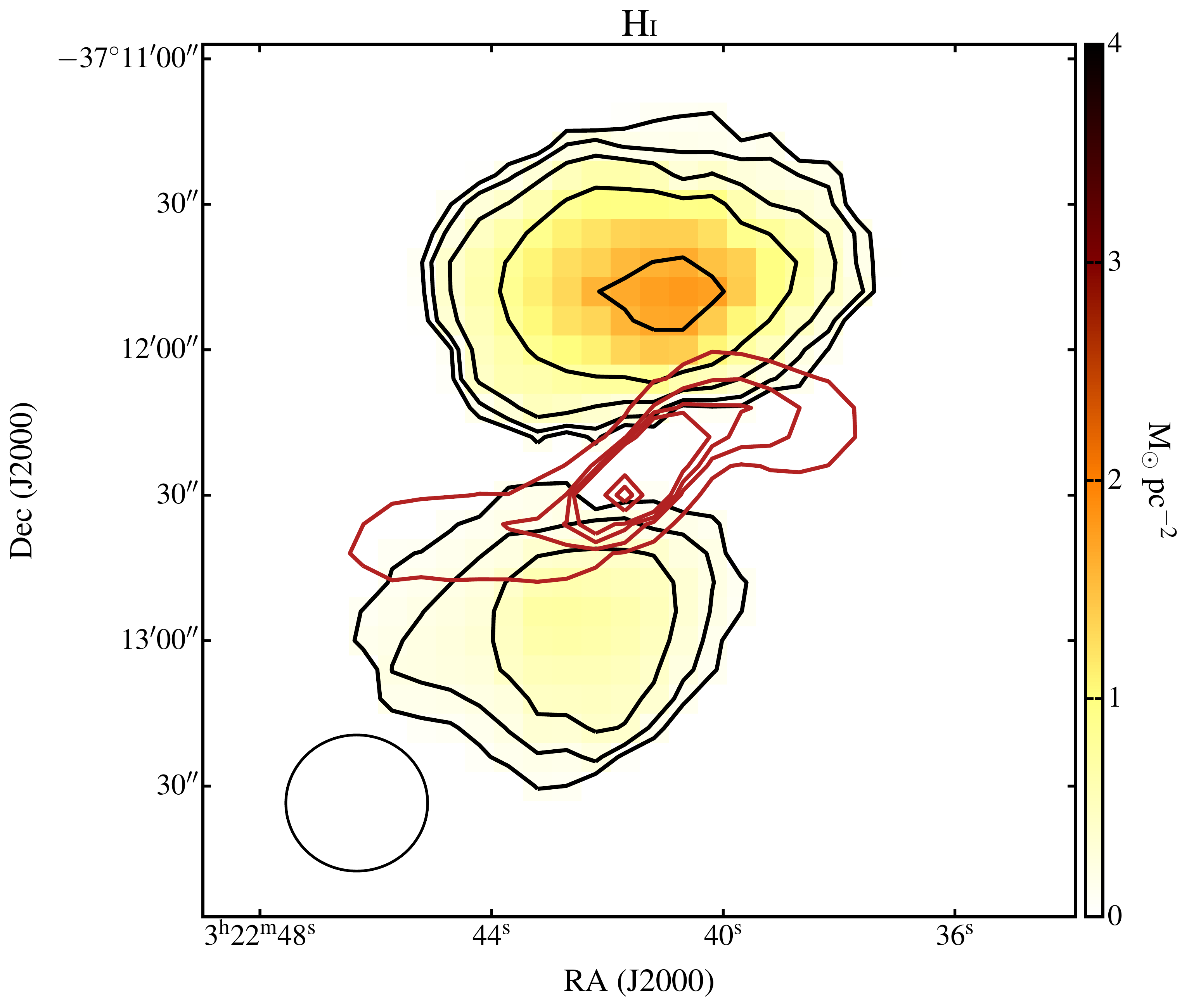}
                \caption{{\em Left panel}: Surface brightness map of the CO 1-0 line detected by ALMA (\cite[Morokuma-Matsui et al. 2019]{morokuma2019}), overlaid with the radio jets. Contour levels are $3\times2^n$~\msun\,pc$^{-2}$,  (n = 0, 1, 2, ...). {\em Right panel}: Surface brightness map of the \HI\ detected by \meer~(\cite[Serra et al. 2019]{serra2019}), overlaid with the radio jets. Contour levels are $0.1\times2^n$~\msun\,pc$^{-2}$,  (n = 0, 1, 2, ...). }
                \label{fig:coldgas} 
        \end{center}
\end{figure} 

\firstsection 

\section{Neutral and molecular gas in the centre of \forn}

The kinematics and distribution of the cold gas in the innermost $6$ kpc (Fig.\,\ref{fig:coldgas}) provide further information on the last episode of the recurrent activity of \forn, which generated the central jets. In the centre, ALMA observations detect $5.6\times 10^8$~\msun\ of molecular hydrogen (\Htwo) distributed in a clumpy shell around the jets~(\cite[Morokuma-Matsui et al. 2019]{morokuma2019}). Neutral hydrogen (\HI) clouds ($4 \times 10^7$~\msun) are closely associated with the molecular gas~(\cite[Serra et al. 2019]{serra2019}). The \HI\ seems to be more extended than the \Htwo, forming a halo where the molecular clouds are embedded. This is confirmed by new, yet unpublished, MeerKAT higher resolution observations which reveal a diffuse \HI\ component surrounding the radio jets. The jets bend in the denser regions of the gas distribution (in proximity of \Htwo\ clouds with irregular kinematics) toward sparser regions (where no molecular clouds are detected but only diffuse \HI, Fig.\,\ref{fig:coldgas} right panel) suggesting a tight interplay between the nuclear activity and the surrounding cold interstellar medium.

\firstsection 




\begin{thebibliography}{}


\bibitem[Fabian (2012)]{fabian2012} {Fabian A. C.} 2012, \textit{ARAA}, 50, 455

\bibitem[Fomalont et al.(1989)]{fomalont1989}{Fomalont, E.~B., Ebneter, K.~A., van Breugel, et al.} 1989, \textit{APJ} (Letters), 346, L17

\bibitem[Galametz et al.(2014)]{galametz2014} {Galametz, M., Albrecht, M., et al.} 2014, \textit{MNRAS}, 439, 2542

\bibitem[Geldzahler \& Fomalont(1984)]{geldzahler1984} {Geldzahler, B.~J., \& Fomalont, E.~B.} 1984,\textit{AJ}, 89, 1650


\bibitem[{Harwood, et al.}{2013}]{harwood2013} {Harwood J.~J., Hardcastle M.~J., Croston J.~H., et al.} 2013, \textit{MNRAS}, 435, 3353

\bibitem[Horellou et al.(2001)]{horellou2001} {Horellou, C., Black, J.~H., van Gorkom, J.~H., et al.} 2001, \textit{A\&A}, 376, 837
\bibitem[Hurley-Walker et al.(2017)]{hurleywalker2017} {Hurley-Walker, N., Callingham, J.~R., Hancock, P.~J., et al.} 2017, \textit{MNRAS}, 464, 1146

\bibitem[Iodice et al.(2017)]{iodice2017} {Iodice, E., Spavone, M., Capaccioli, M., et al.} 2017, \textit{ApJ}, 839, 21

\bibitem[Jonas \& MeerKAT Team(2016)]{jonas2016} Jonas, J., \& MeerKAT Team\ 2016, \textit{Proceedings of MeerKAT Science}, 1

\bibitem[Kardashev (1962)]{kardashev1962} {Kardashev, N.~S.} 1962, \textit{Soviet Astron.}, 6, 317


\bibitem[{Kolokythas, et al.}{2015}]{kolokythas2015} {Kolokythas K., O'Sullivan E., Giacintucci S., et al.} 2015,\textit{MNRAS}, 450, 1732

\bibitem[Lanz et al.(2010)]{lanz2010}{Lanz, L., Jones, C., Forman, W.~R., et al.} 2010, \textit{ApJ}, 721, 1702


\bibitem[{Maccagni, et al.}{2020}]{maccagni2020} {Maccagni F.~M., Murgia, M., Serra, P., et al.} 2020, \textit{A\&A}, 634, A9

\bibitem[McKinley et al.(2015)]{mckinley2015} {McKinley, B., Yang, R., L{\'o}pez-Caniego, M., et al.} 2015, \textit{MNRAS}, 446, 3478

\bibitem[Morganti(2017)]{morganti2017} {Morganti, R.} 2017, \textit{Nature Astronomy}, 1, 596


\bibitem[{Morokuma-Matsui, et al.}{2019}]{morokuma2019} {Morokuma-Matsui K., Serra P., Maccagni, F. M., et al.} 2019, \textit{PASJ}, 71, 85


\bibitem[Murgia, et al. 1999]{murgia1999} {Murgia M., Fanti C., Fanti R., et al.} 1999, \textit{A\&A}, 345, 769


\bibitem[{Murgia, et al.}{2011}]{murgia2011} {Murgia M., Parma, P. Mack, H. K., et al.} 2011, \textit{A\&A}, 526, A148


\bibitem[Planck Collaboration et al.(2018)]{planck2018} {Planck Collaboration, Akrami, Y., Ashdown, M., et al.} 2018, \textit{arXiv e-prints}, arXiv:1807.06208


\bibitem[Prandoni et al.(2017)]{prandoni2017} {Prandoni, I., Murgia, M., Tarchi, A., et al.} 2017, \textit{A\&A}, 608, A40

\bibitem[Ramos Almeida et al.(2012)]{ramosalmeida2012} {Ramos Almeida, C., Bessiere, P.~S., Tadhunter, C.~N., et al.} 2012, \textit{MNRAS}, 419, 687

\bibitem[Sabater et al.(2013)]{sabater2013} {Sabater, J., Best, P.~N., \& Argudo-Fern{\'a}ndez, M.} 2013, \textit{MNRAS}, 430, 638

\bibitem[Schawinski et al.(2015)]{schawinski2015} {Schawinski, K., Koss, M., Berney, S., et al.} 2015, \textit{MNRAS}, 451, 2517


\bibitem[{Serra, et al.}{2019}]{serra2019} {Serra P., Maccagni, F. M., Kleiner, D., et al.} 2019, \textit{A\&A}, 628, A122

\bibitem[{Sesto, et al.}{2018}]{sesto2018} {Sesto L.~A., Faifer F.~R., Smith Castelli A.~V., et al.} 2018, \textit{MNRAS}, 479, 478

\bibitem[Tashiro et al.(2009)]{tashiro2009} {Tashiro, M.~S., Isobe, N., Seta, H., et al.} 2009, \textit{PASJ}, 61, S327

\bibitem[Werner, et al. (2019)]{werner2019} {Werner N., McNamara B.~R., Churazov E., et al.} 2019, \textit{SSRv}, 215, 5

\end{thebibliography}
\end{document}